# $A^3Rank$: Augmentation Alignment Analysis for Prioritizing Overconfident Failing Samples for Deep Learning Models

ZHENGYUAN WEI, City University of Hong Kong & The University of Hong Kong, Hong Kong

HAIPENG WANG, City University of Hong Kong, Hong Kong

QILIN ZHOU, City University of Hong Kong, Hong Kong

W.K. CHAN, City University of Hong Kong, Hong Kong

Sharpening deep learning models by training them with examples close to the decision boundary is a well-known best practice. Nonetheless, these models are still error-prone in producing predictions. In practice, the inference of the deep learning models in many application systems is guarded by a rejector, such as a confidence-based rejector, to filter out samples with insufficient prediction confidence. Such confidence-based rejectors cannot effectively guard against failing samples with high confidence. Existing test case prioritization techniques effectively distinguish confusing samples from confident samples to identify failing samples among the confusing ones, yet prioritizing the failing ones high among many confident ones is challenging. In this paper, we propose $A^3Rank$, a novel test case prioritization technique with augmentation alignment analysis, to address this problem. $A^3Rank$ generates augmented versions of each test case and assesses the extent of the prediction result for the test case misaligned with these of the augmented versions and vice versa. Our experiment shows that $A^3Rank$ can effectively rank failing samples escaping from the checking of confidence-based rejectors, which significantly outperforms the peer techniques by 163.63% in the detection ratio of top-ranked samples. We also provide a framework to construct a detector devoted to augmenting these rejectors to defend these failing samples, and our detector can achieve a significantly higher defense success rate.

## 1 INTRODUCTION

Although deep learning (DL) models perform acceptably well in evaluation, these DL models still inevitably produce wrong predictions in practice. Wrong predictions can lead to various problems in different application domains, e.g., improper medical diagnosis [25] and traffic accidents [16]. Enhancing the DL application systems by reducing wrong predictions of DL models in producing outputs is desirable. Studies [9, 51, 52] have shown that DL models are vulnerable to operational input samples that can lead them to produce incorrect predictions in natural scenarios [52], and the prediction confidences of many such failing samples exceed those well-intended guarding confidence levels [54]. For example, strong sunshine may cause the camera of a self-driving car to capture an image full of white pixels, resulting in a prediction failure with high confidence. A major bottleneck in developing DL applications is detecting these overconfident failures from their deployed DL application systems.

To reduce unreliable predictions, many real-world machine-learning-based application systems are equipped with rejectors to discard uncertain decisions [17]. In DL application systems, many existing techniques [6, 17, 45] construct their rejectors for DL models to address the incorrect prediction problem. For example, many recent studies [2, 8, 42, 49] have been conducted to enhance the defense ability of DL models against out-of-distribution (OOD) samples from unknown classes or artificial examples that are very likely to guide DL models to yield failures. On the other hand, for the failures caused by in-distribution (ID) samples from known classes, existing techniques tend to detect those failing samples close to the decision boundaries of DL models [11, 32, 46]. Detecting the failures of in-distribution samples that have non-trivial distances away from the decision boundaries of DL models is under-explored.

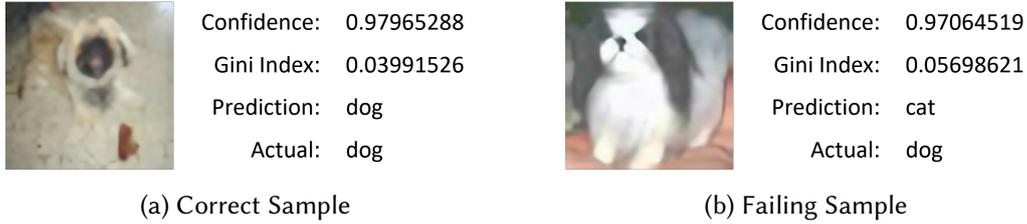

(a) Correct Sample          (b) Failing Sample

Fig. 1. A pair of natural images from [18] with over 0.9 prediction confidence evaluated on a pre-trained model [5]. The state-of-the-art prioritization technique [11] still fails to separate them for better detection.

A widely-used and effective solution [20] is to exploit the maximum softmax probability from a classifier (aka. prediction confidence) to filter out (e.g., reject) an input sample with the prediction probability of the prediction class not much larger than those of the remaining classes. Another well-known solution is DeepGini [11] which measures the gini impurity (a kind of measurement of uncertainty) from the probability vector produced by a DL model, and lower prediction confidence in the probability vector implies a higher uncertainty. Previous studies reported that some failing examples (e.g., an example with overly high prediction confidence) were not easily detected [4], and the existing state-of-the-art rejectors are still far from the perfect defense success rate [1]. Fig. 1 shows a real-world example revealing this situation. As such, the failing samples that can escape from rejectors still silently threaten the DL models. It is desirable to develop DL testing techniques to identify these escaping failing samples for follow-up analysis, such as developing a stronger rejection to augment the existing rejectors and filter these failing samples.

In the context of DL testing, developers often focus on the tests that lead to prediction failures for the reason that analyzing these tests can provide insights to safeguard the DL systems. However, unlike conventional software with human-developed business logic for generating tests, DL testing generally faces automated testing oracle problems. This fact motivates us to propose a testing technique to prioritize tests so that fault-inducing tests can be labeled and analyzed before the other tests. In this manner, we can obtain maximum benefit from human efforts and leverage fault-inducing tests to construct or augment the rejectors.

In this paper, we propose $A^3Rank$ (**A**ugmentation **A**lignment **A**nalysis for **Rank**ing), a novel and effective test case prioritization (TCP) technique to prioritize confident failing samples of DL models escaped from rejectors. $A^3Rank$ tackles the challenge that failing samples escaping from rejectors are more difficult to find. For instance, the proportion of these samples in all failing samples becomes smaller as the prediction confidence threshold for rejection increases. We focus on the classification problem as a starting point, as it is one of the fundamental prediction problems many TCP techniques in DL model testing concentrate on. We observe that relying on a single prediction vector of the sample under test could be unreliable in ranking failing samples with high confidence levels, as the DL models under test have shown confidence in predicting their labels. On the other hand, we observe that DL models in real-world DL applications are data-augmented. Thus, $A^3Rank$ proposes to perform augmentation-based mutation analysis on the alignment among multiple prediction vectors of the test sample and its mutated samples.

The experimental results show that $A^3Rank$ significantly outperforms the baselines (DeepGini, Dropout, Dissector, LSA, and Prima) averagely by 163.63% on the detection ratio of failing samples that escaped from confidence-based rejectors, and the defender constructed from the failing samples ranked high by $A^3Rank$ achieves a higher defense success rate of 12.67% than the defenders produced through the peer TCP techniques.

The main contribution of this paper is threefold: (1) This paper presents *the first work* to tackle the challenge of effectively prioritizing (over)confident failing samples that escape from confidence-based rejectors. (2) It proposes $A^3Rank$ technique with its novel augmentation alignment analysis. (3) It shows the effectiveness and feasibility of $A^3Rank$ framework of its two-stage rejection strategy.

The rest of the paper is organized as follows. Section 2 revisits the preliminaries. Section 3 revisits the rejection in DL models. Sections 4 and 5 present $A^3Rank$ and its evaluation. We review the related works in Section 6 and conclude the paper in Section 7.

## 2 PRELIMINARIES

### 2.1 Classification Models

A classification model (classifier for short) M with $C$ class labels $\{c_1, \cdots, c_C\}$ is a function. It takes a sample $s$ as input and outputs a *prediction vector*, denoted as $M(s)$. The number of components of $M(s)$ is $C$. The $i$-th component of $M(s)$ (for $1 \leq i \leq C$) keeps a value, denoted by $M(s)_i$, representing the probability of $s$ predicted as class $c_i$. The summation of the values in $M(s)$ is 1, i.e., $\sum_{i=1}^{C} M(s)_i = 1$. The prediction class of $s$, denoted by $c_p$, is the class for the component of $M(s)$ keeping the largest value, i.e., $p = \arg\max_{1 \leq i \leq C} M(s)_i$. The value $M(s)_p$ is called the *prediction confidence* of $s$. Each sample $s$ after labeling associates with a groundtruth label $l_s$. If $c_p \neq l_s$ (i.e., the prediction class differs from the groundtruth label), $s$ is called a *failing sample* or *failing test case*. If $c_p = l_s$, M is said to correctly classify $s$, otherwise misclassify it.

### 2.2 Models with Data Augmentation Training

Training samples in many application domains (e.g., computer vision) are popularly data-augmented [43] when used to train a classifier M. For example, a training sample can be augmented by a set of geometric operations (e.g., shifting) in the image classification task. Let $A$ be a set of transformation operations to transform a sample $s$. Applying each operation in $A$ transforms $s$ into a *sample variant* $\tilde{s}$ of $s$ paired with the groundtruth label $l_s$ of $s$ if available. We refer to the set of all sample variants of $s$ after applying $A$ as $\mathcal{S}$. Many studies [43? ? ] show that data augmentation can improve the test accuracy and robustness of DL models. Therefore, many practical DL models are produced with training schemes using data-augmented training sets.

### 2.3 Classification Model with Reject Option

A classifier may misclassify some input samples. An application system deployed with such a classifier is often equipped with a "guard" to filter out some samples that tend to be misclassified by the classifier. A recent survey [17] reviews machine learning models with a reject option, where a model with a reject option, H, consists of a classifier M and a rejector R. The survey also summarizes that most existing works it has reviewed take the prediction confidence of M as a metric to define a rejector. For instance, if the prediction confidence for a sample $s$ made by M is lower than 0.7, then R outputs true, otherwise false. More specifically, in addition to all $C$ classes of the classifier M, the model H has one more class $c_\times$, representing the class of samples to be rejected even if the sample

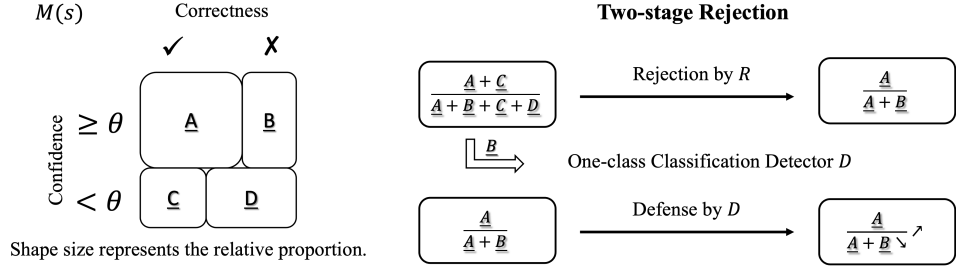

Fig. 2. Two-stage rejection of $A^3 Rank$ framework.

has been predicted by M. Formally, a model with a reject option, H = ⟨M, R⟩, is defined as follows:

$$H(s) = \begin{cases} c_\times & \text{if } R(s, M) \text{ is true} \\ M(s) & \text{otherwise} \end{cases} \tag{1}$$

Suppose $T$ is a set of samples (captured under the natural scenarios) for the model H to classify. In general, some samples in $T$ are failing samples of the classifier M in H, and some failing samples may also pass the rejector R (i.e., R($s'$, M) is false for some failing sample $s'$ of M). We refer to a failing sample of M that passes the rejector R as a ***subtle sample*** in the context of the model H.

## 3 TWO-STAGE REJECTION

Suppose H = ⟨M, R⟩ is a classifier with a reject option, in which M is an accurate data-augmented classifier under test and R is a rejector that rejects an input sample $s$ (i.e., returns true) if the prediction confidence of M on $s$ is lower than a predefined threshold value. Suppose further $T$ is a set of unlabeled samples input to H.

We aim to prioritize $T$ so that the failing samples of M in $T$ are ranked as high as possible, subject to the condition that the rejector R does not reject them. In other words, we want to find these unlabeled failing samples predicted by M that can pass through R. The samples ranked higher in the prioritized $T$ are then labeled to reveal whether they are failing samples. A one-class classification [36] (OCC)-based detector D using these labeled and subtle samples is created to extend the original reject option of H to check each input sample (that passes R) by D, thereby creating a modified model H′ = ⟨M, R′⟩ of H where R′ = ⟨R, D⟩.

$$H'(s) = \begin{cases} c_\times & \text{if } R(s, M) \vee D(s, M) \text{ is true} \\ M(s) & \text{otherwise} \end{cases} \tag{2}$$

Fig. 2 presents the benefit of such a two-stage rejection scheme. The predictions of model $M$ can be divided into four cases in two dimensions, i.e., whether $M$ predicts correctly and whether the prediction confidence exceeds a confidence threshold $\theta$. Namely, the four cases are denoted by $\underline{A}$, $\underline{B}$, $\underline{C}$, and $\underline{D}$, respectively, according to the dimensions in the figure. The sizes of their shapes depict the intuition on the relative proportion of samples among these four cases for typical well-trained DL models (e.g., the pretrained models on ImageNet). After rejection by $R$, the model accuracy of H over the remaining inputs is $\underline{A}/(\underline{A}+\underline{B})$ (note that the model accuracy of M is $(\underline{A}+\underline{C})/(\underline{A}+\underline{B}+\underline{C}+\underline{D})$). Although case $\underline{D}$ is eliminated, case $\underline{B}$ remains in the equation for model H. To further improve the model accuracy in the equation of $\underline{A}/(\underline{A}+\underline{B})$, a popular approach is to defend the novel samples of $\underline{B}$ based on the identified samples of $\underline{B}$ with an OCC-based detector $D$. Therefore, the model H′ contains $D$ to filter those subtle samples, thereby increasing its model accuracy in the equation of $\underline{A}/(\underline{A}+\underline{B})$,

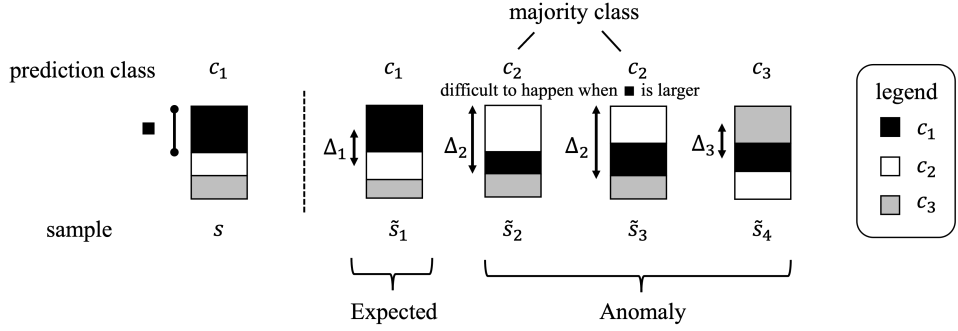

Fig. 3. Anomaly pattern of majority misalignment poses suspiciousness on model predictions

For a model with a reject option $H = \langle M, R \rangle$, subtle samples, which are also failing samples, are interesting because they not only trigger the classifier $M$'s failures but also escape from the guard of the rejector $R$. Finding them earlier for labeling and improving $H$ is thus attractive.

However, we have to overcome a challenge: Suppose $T$ is a set of unlabeled samples. For an accurate classifier $M$, many samples will be classified with high confidence (e.g., 0.9706 as illustrated in Fig. 1(b)). However, as the prediction confidence increases, we observe that the ratio of failing samples tends to drop. In this case, ranking failing samples high becomes increasingly difficult. Suppose two test samples $s$ and $s'$ have almost the same prediction confidence (or prediction vectors) in a high prediction confidence range (e.g., able to pass the rejector $R$ with a large margin), and only the test sample $s$ is a failing one. Before labeling them, it is unclear how to effectively decide whether $s$ is more likely to lead $M$ to misbehave than $s'$. In our experiment (Section 5), we will show that the current state-of-the-art rankers (e.g., DeepGini [11]) are inadequate.

## 4 OUR TECHNIQUE

### 4.1 Motivation

Inspired by the ineffective analysis result depicted in Fig. 2, relying merely on single prediction confidence appears ineffective in detecting the failing samples in case $\underline{B}$, which motivates us to explore new ideas to test samples from the reliability perspective. We observe that the augmented samples (called sample variants) produced through perturbating a training sample are all trained to be predicted to the same groundtruth label during augmentation training, which provides a clue for testing that the prediction classes of these sample variants of a test sample are expected to match the prediction class of the test sample itself. This insight is also consistent with our experience that a failure (i.e., misclassification) would generally occur when the majority of the prediction classes of these sample variants of a test sample is different from the prediction class of the test samples.

Fig. 3 illustrates this anomaly pattern. Suppose a test sample $s$ is classified to class $c_1$, and its sample variants $\{\tilde{s}_1, \tilde{s}_2, \tilde{s}_3, \tilde{s}_4\}$ are predicted to classes $c_1$, $c_2$, $c_2$, and $c_3$, respectively, by a 3-class classifier. Producing anomaly predictions by sample variants, especially the ones predicted to the majority prediction class (i.e., $c_2$ in this example), tends to be more difficult if the test sample $s$ has higher and higher prediction confidence. The difference in the prediction probabilities between sample variants and the test sample reveals a clue on misalignment. This motivates us to formulate a comprehensive and quantitative analysis to measure the misalignment through prediction differences among the sample test and its variants from the reliability angle.

## 4.2 Augmentation Alignment Analysis

We suppose the classifier under test M is trained with data augmentation. In the training scheme to produce M, the training set is augmented by applying a set of data augmentation transformation operations $A$ to each training sample.

Let $s$ be a test sample to be ranked. We apply each operation in $A$ to transform $s$ into a sample variant. We denote the set of all generated sample variants of $s$ by $\mathcal{S}$.

To address the abovementioned challenge, we propose to exploit the difference between a sample and its data-augmented versions (i.e., *sample variants*). We aim to explore both alignment and misalignment between how confident M classifies the sample $s$ to its prediction class but not to the classes of its sample variant set $\mathcal{S}$ and how confident the same classifier classifies individual sample variants to their prediction classes but not to the prediction class of the test sample $s$.

We first clarify how to classify the set $\mathcal{S}$, and the formal definition can be found in the later part of this subsection. The classifier M will classify each sample variant in $\mathcal{S}$ into a class. The (majority) class, denoted by $c_m$, associated with the largest counts among sample variants, is treated as the prediction class of $\mathcal{S}$. Furthermore, to simplify our presentation, we refer to the prediction class of the test sample $s$ as $c_p$ (i.e., M classifies $s$ to $c_p$).

In this work, we propose $A^3Rank$, a technique with the novel **A**ugmentation **A**lignment **A**nalysis for **Rank**ing samples of DL models with reject options. $A^3Rank$ connects each test sample $s$ and individual sample variants not only directly by their prediction classes (and confidence) but also indirectly via the prediction class of $\mathcal{S}$ from both alignment and misalignment angles. We aim to provide an analysis suitable for all samples under test. As such, we design the $A^3Rank$ formula applicable for ranking confident failing samples and confusing failing samples.

We refer to the abovementioned class $c_m$ as the *majority prediction class* of the sample $s$. Given an input sample $s$ and its sample variant set $\mathcal{S}$, this majority prediction class $c_m$ is computed by finding the class index $m$ using the following formula: $m = \arg\max_{1 \leq j \leq C} \sum_{\tilde{s} \in \mathcal{S}} \mathbb{1}\{\arg\max_{1 \leq i \leq C} \mathsf{M}(\tilde{s})_i = j\}$ where $\mathbb{1}\{J\}$ is a function that returns 1 if the condition $J$ holds, otherwise 0. A tie case is resolved by choosing the class with higher mean prediction confidence.

We further distinguish each sample variant's roles in our differential alignment analysis.

(1) Dominator ($\tilde{s}_{dom}$). If the classifier M classifies a variant $\tilde{s}_{dom}$ to the majority prediction class $c_m$ (i.e., $m = \arg\max_i \mathsf{M}(\tilde{s}_{dom})_i$), the variant $\tilde{s}_{dom}$ is called a dominator. This role indicates the variant contributes to converging the decision of the prediction class for $\mathcal{S}$.

(2) Supporter ($\tilde{s}_{sup}$). If M classifies $\tilde{s}_{sup}$ to the prediction class $c_p$ of the sample $s$ (i.e., $p = \arg\max_i \mathsf{M}(\tilde{s}_{sup})_i$), the variant $\tilde{s}$ is called a supporter. This role indicates that the variant aligns with the sample under rank $s$ in their prediction results. We note that a sample variant may be simultaneously a dominator and a supporter when $c_p = c_m$.

(3) Distractor ($\tilde{s}_{dis}$). If M classifies a variant $\tilde{s}_{dis}$ to a class other than both $c_p$ and $c_m$, then the variant $\tilde{s}_{dis}$ is called a distractor. The variant serves as a distractor to discourage a consistent decision on the prediction classes between $s$ and the set $\mathcal{S}$.

Moreover, we denote the sets of all dominators, all supporters, and all distractors by $\mathcal{S}_{dom}$, $\mathcal{S}_{sup}$, and $\mathcal{S}_{dis}$, respectively. A sample variant is both a supporter and a dominator if the majority prediction class $c_m$ is the same as the prediction class $c_p$.

The prediction confidence presents the likelihood of the prediction reliability of a sample. However, since a sample with higher prediction confidence may or may not be failing, directly adopting the prediction confidence of a sample as the ranking metric is not adequate. Therefore, $A^3Rank$ uses the prediction confidence of a sample to rank as the baseline and formulates a novel suite of three alignments on this baseline to assess the sample prediction reliability for ranking them.

*4.2.1 Alignment from Distractors.* The presence of a distractor $\tilde{s}_{dis}$ indicates that the prediction on the test sample $s$ by the classifier M encounters a certain extent of "hidden" uncertainty that can be observed "around" the test sample $s$. Our insight is that if the extent of such uncertainty exhibited by a distractor can be resolved, then the distractor will become a supporter, strengthening the reliability of the sample $s$ being correct.

Thus, we design the alignment term $g_1$ to account for the presence of a distractor $\tilde{s}_{dis}$, which is the amount of increase in prediction probability of the class $c_p$ in the prediction vector of $\tilde{s}_{dis}$ so that the class $c_p$ could emerge into the prediction class of this distractor $\tilde{s}_{dis}$.

$$g_1(\tilde{s}_{dis}) = \max_{1 \leq i \leq C} M(\tilde{s}_{dis})_i - M(\tilde{s}_{dis})_p \tag{3}$$

*4.2.2 Alignment from Supporters.* On the contrary, each supporter $\tilde{s}_{sup}$ enhances the prediction reliability on the test sample $s$. Recall that the prediction by an ensemble over a set of classifiers that returns the majority output is more robust than the prediction of individual classifiers in the ensemble. Suppose $c_p = c_m$ in the prediction vector of this supporter $\tilde{s}_{sup}$. In this case, the supporter shows robust support for the prediction class of the test sample $s$. On the other hand, suppose $c_p \neq c_m$ for $\tilde{s}_{sup}$, the supporter provides an (over)confidence over the more robust the majority label to support the test sample $s$. To provide reliable support to the test sample $s$, the prediction probability for the majority class in the prediction vector of $\tilde{s}_{sup}$ should be at least increased to the same level as the prediction probability for the class $c_p$ in that prediction vector. Thus, we design the following alignment term $g_2$ for $\tilde{s}_{sup}$:

$$g_2(\tilde{s}_{sup}) = M(\tilde{s}_{sup})_p - M(\tilde{s}_{sup})_m \tag{4}$$

*4.2.3 Alignment from Dominators.* The dominators may show strong uncertainty or great support for the prediction reliability of the test sample $s$, depending on whether the majority prediction class $c_m$ differs from the prediction class $c_p$ of the sample $s$. More specifically, if $c_m = c_p$, the dominators highly support $s$ to be classified to $c_p$ (and thus, $s$ is less likely to be a failing sample intuitively). Nonetheless, if $c_m \neq c_p$, the presence of dominators indicates a large extent of uncertainty that $s$ belongs to the class $c_p$. Intuitively, the latter case ($c_m \neq c_p$) occurs less frequently than the former ($c_m = c_p$) due to the application of data augmentation to push a sample and its sample variants generally toward the same label. Moreover, when a scenario in the latter case occurs, as the ratio of dominators to all the variants in $\mathcal{S}$ increases, the test sample $s$ is more likely to be a failing sample.

Owing to the need to compare the test sample $s$ and its sample variants that dominate in the set $\mathcal{S}$, the extent of misalignment in prediction reliability on $s$ has two sources. The first source is the loss in prediction probability needed to align a dominator $\tilde{s}_{dom}$ to become a supporter. The larger the loss is, the more confused M is to distinguish $s$ between $c_m$ and $c_p$. The second one is the loss in prediction probability needed to change the test sample $s$ from the current prediction class $c_p$ to the majority prediction class $c_m$ of its variant set so that $s$ can align with the set $\mathcal{S}$ in prediction. The larger the loss implies the larger over-confidence of M to classify $s$ to $c_p$ relative to $c_m$. If both losses can be resolved, a dominator will greatly support the reliability that $s$ is predicted correctly.

We present the alignment term $g_3$ for a dominator $\tilde{s}_{dom}$ from the reliability gain perspective. We note that the evaluation results of the alignment terms of dominators may be negative values.

$$g_3(\tilde{s}_{dom}) = [M(\tilde{s}_{dom})_p - M(\tilde{s}_{dom})_m] + [M(s)_m - M(s)_p] \tag{5}$$

*4.2.4 Overall Formula.* For each variant in $\mathcal{S}$, $A^3$Rank applies the alignment terms based on their roles as presented above. (Note that a variant may have multiple roles.) The alignment from the distractor explicitly shows a negative effect on the prediction reliability. Thus, $A^3$Rank uses the negation of this alignment term. The augmentation alignment analysis score ($A^3$-score) for a sample $s$ is computed by the following formula $A^3$Score($s$). We linearly combine all the terms because we do not find a good reason to weigh a value more. Test samples are ranked in ascending order of $A^3$-score, so intuitively, less reliable test samples will be prioritized first.

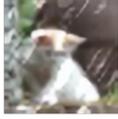

| $c_p = c_3$<br>bird | $c_1$ | $c_2$ | $c_3$ | $c_4$ | $c_5$ | $c_6$ | $c_7$ | $c_8$ | $c_9$ | $c_{10}$ | |
|---|---|---|---|---|---|---|---|---|---|---|---|
| | 0.00 | 0.00 | **0.95** | <u>0.02</u> | 0.01 | 0.01 | 0.01 | 0.00 | 0.00 | 0.00 | $A^3(s) = -1.19$ |
| majority<br>$c_m = c_4$<br>cat | 0.00 | 0.00 | <u>0.27</u> | 0.28 | 0.04 | **0.36** | 0.04 | 0.01 | 0.00 | 0.00 | $g_1(\tilde{s}_{dis}) = 0.09$ |
| | 0.00 | 0.00 | **0.66** | <u>0.06</u> | 0.12 | 0.05 | 0.11 | 0.00 | 0.00 | 0.00 | $g_2(\tilde{s}_{sup}) = 0.60$ |
| | 0.00 | 0.00 | <u>0.05</u> | **0.40** | 0.12 | 0.35 | 0.07 | 0.01 | 0.00 | 0.00 | $g_3(\tilde{s}_{dom}) = -1.28$ |
| | 0.00 | 0.00 | <u>0.16</u> | **0.60** | 0.01 | 0.20 | 0.03 | 0.00 | 0.00 | 0.00 | $g_3(\tilde{s}_{dom}) = -1.37$ |

Fig. 4. An example of the $A^3Rank$ alignment in the experiment. Five prediction vectors of a sample $s$ and its variants $\mathcal{S}$ (from top to bottom) are shown, where each vector has probabilities for ten classes ($c_1 - c_{10}$), and the confidences of the prediction classes are bold. The majority class $c_m$ among the four variants is $c_4$. Their prediction classes are $c_6$, $c_3$, $c_4$, and $c_4$. The four variants are distractor, supporter, dominator, and dominator. We underline the measurement points on each vector for the corresponding alignments (on the right-hand side of that vector). $A^3Rank$ scores the test sample $s$ with a $A^3$-score lower than its prediction class's prediction probability (0.95).

$$A^3\text{Score}(s) = M(s)_p - \sum_{\tilde{s} \in \mathcal{S}_{dis}} g_1(\tilde{s}) + \sum_{\tilde{s} \in \mathcal{S}_{sup}} g_2(\tilde{s}) + \sum_{\tilde{s} \in \mathcal{S}_{dom}} g_3(\tilde{s}) \qquad (6)$$

*4.2.5 Example.* Fig. 4 illustrates an example to carry out the analysis of $A^3Rank$ in the experiment. The sample $s$ is predicted by M with a confidence of 0.95. $A^3Rank$ generates its sample variants, and four of them are illustrated with the prediction vectors below the one for the sample $s$. However, the analysis of $A^3Rank$ shows a large misalignment from the sample variants of $s$. The distractor $\tilde{s}_{dis}$ gives 0.09 in the strength of uncertainty for the prediction, and the supporter $\tilde{s}_{sup}$ supports the prediction with 0.60 in strength. However, the two dominators agree to the consensus of class $c_4$ and show 1.28 and 1.37 in strengths of the misalignment, respectively. Thus, $A^3Rank$ scores $s$ according to Eq. 6 to $0.95 - 0.09 + 0.60 - 1.28 - 1.37 = -1.19$.

## 4.3 Labeling Samples

The ranked list of samples is presented to developers. They then label samples, and thus, failing samples are revealed. After that, $A^3Rank$ proceeds to improve the model $H$ from the detection perspective, presented in the next subsection.

## 4.4 Enhancing the Model with Reject Option with $A^3$Rank

$A^3Rank$ applies the discovered subtle samples to construct an OCC detector $D$ to defend further the samples that pass the rejector R. Specifically, the set of discovered subtle samples is viewed as a single class of samples to construct $D$. Thus, any sample predicted to be within the distribution by the trained $D$ is deemed failing and thus rejected, as summarized in Eq. 2.

In our experiment, we adopt NIC [31] to generate our OCC detector on account of its superior effectiveness stated in their paper [31] and comprehensive profiling of internal states of M. The original NIC algorithm accepts the clean training set of M to produce an OCC detector to exclude samples described by the clean training set. We input the abovementioned set of discovered and labeled subtle samples to the NIC algorithm to return an OCC detector D. In the inference time, the OCC detector D checks whether a novel sample is similar to the learned knowledge from the set of

---

**Algorithm 1:** $A^3$Rank Framework

| | |
|---|---|
| **Input** | :Model H = ⟨ Classifier M, Rejector R ⟩, |
| | Natural samples $T_{nat}$, Labeling resource $\omega$ |
| **Output** | :Enhanced model H′ |
| | Ranked list of samples $T_{nat}^{\triangleleft}$ |

**1** $L \leftarrow$ empty list
**2** $A \leftarrow$ RetrieveAugmentationOperators(M)
**3** **foreach** $s \in T_{nat}$ **do**
**4**     $c_p, v_p \leftarrow$ Predict(M, $s$)
**5**     $\mathcal{S} \leftarrow$ Augment($s$, $A$)
**6**     $\mathbf{c}, \mathbf{v} \leftarrow$ Predict(M, $\mathcal{S}$)
**7**     $c_m \leftarrow$ Majority($\mathbf{c}, \mathbf{v}$)
**8**     $L \Leftarrow A^3$Score($c_p, v_p, c_m, \mathbf{c}, \mathbf{v}$)
**9** **end foreach**
**10** $T_{nat}^{\triangleleft} \leftarrow$ Rank(M, $T_{nat}$, $L$)
**11** $T_{sub} \leftarrow$ Label($T_{nat}^{\triangleleft}$, R, $\omega$)
**12** D $\leftarrow$ ConstructDetector($T_{sub}$)
**13** R′ = ⟨R, D⟩
**14** H′ = ⟨M, R′⟩
**15** **return** H′, $T_{nat}^{\triangleleft}$

---

labeled subtle samples. If this is the case, the detector returns true, otherwise false. We note that since $A^3Rank$ intends to deal with the subtle samples, the trained D only checks the input samples that have passed the original rejector R.

Alg. 1 summarizes the whole $A^3Rank$ framework. It accepts a model H consisting of a data-augmented classifier M and a rejector R, a set of natural samples $T_{nat}$, and the label resource $\omega$. Line 1 initializes an empty list $L$, and line 2 retrieves the set of transformation operations $A$ of the augmentation specified in M's training scheme.

The algorithm computes the $A^3$-score for each sample $s$ in $T_{nat}$ in the loop (lines 3–9). It applies M to classify $s$ to produce the prediction class $c_p$ and the prediction vector $v_p$ (line 4). It then generates a set of sample variants $\mathcal{S}$ using $A$ on $s$ (line 5), followed by applying M to produce their corresponding prediction classes $\mathbf{c}$ and prediction vectors $\mathbf{v}$ (line 6), both are in the vector form. It finds the majority prediction class $c_m$ of $\mathcal{S}$ (line 7). In line 8, it computes the $A^3$-score according to Eq. 6, and then adds the pair of the sample and its $A^3$-score to $L$. The algorithm ranks the samples in $T_{nat}$ by their $A^3$-scores kept in $L$ to produce a ranked list of samples $T_{nat}^{\triangleleft}$ (line 10), in which samples are sorted in ascending order of their $A^3$-scores. The next step is to construct a sublist consisting of the top-$\omega$ samples of $T_{nat}^{\triangleleft}$, and drop every sample $s$ in the sublist that the rejector R($s$, M) returns true (i.e., rejecting the sample), followed by an external (probably manual) process of sample labeling on the reduced sublist, resulting in a subset of subtle samples with labels $T_{sub}$ (line 11). Next, it constructs an OCC detector $D$ using the subset $T_{sub}$ as its training set (line 12) and then enhances the model H with $D$ (lines 13–14).

## 5 EVALUATION

This section reports the evaluation on $A^3Rank$.

Table 1. Descriptive Statistics of Samples under Rank

| Dataset | #Samples | Failing Ratio | Subtle Ratio | | |
| --- | --- | --- | --- | --- | --- |
| | | | $\theta$=0.7 | $\theta$=0.8 | $\theta$=0.9 |
| CIFAR10 | 50,000 | 14.70% | 9.63% | 8.00% | 5.80% |
| CIFAR100 | 50,000 | 38.05% | 17.88% | 13.46% | 8.85% |
| CINIC10 | 90,000 | 18.01% | 9.32% | 6.89% | 4.16% |

Failing Ratio = #Failing Samples ÷ #Samples
Subtle Ratio = #Subtle Samples ÷ #Samples with $> \theta$ confidence

### 5.1 Research Questions

We aim to answer the following four research questions,

- RQ1: How effective is $A^3Rank$ in detecting subtle samples compared to the baselines?
- RQ2: How effective can $A^3Rank$ framework be in defending subtle samples?
- RQ3: How large improvement does $A^3Rank$ improve in prioritizing failing samples compared with the state-of-the-art techniques?
- RQ4: How do the alignment terms in Eq. 6 contribute to the effectiveness of $A^3Rank$?

### 5.2 Implementation

We implement all the experiments in Python v3.8 and PyTorch v1.12.1 with Torchvision v0.13.1. We run all the experiments on a Ubuntu 20.04 server with a 48-core 3.0GHz Xeon CPU, 256GB RAM, and 2080Ti GPU. We adopt the open-source implementations of the baseline techniques and adapt them to our platform if available. If unavailable, we implement them according to their papers.

### 5.3 Datasets and Models

We aim to evaluate $A^3Rank$ with a concentration on natural scenarios. Therefore, we select three evaluation datasets containing color images with natural classes widely used in DL model testing.

- CIFAR10 [27]: It consists of 60,000 32x32 color images in 10 classes, with 6,000 images per class. There are 50,000 training samples and 10,000 test samples.
- CIFAR100 [27]: The data statistic of CIFAR100 is the same as CIFAR10, except that it has 100 classes containing 600 images per class.
- CINIC10 [7]: It consists of 270,000 32x32 color images in 10 classes, with 27,000 images per class. It has equally sized train, validation, and test splits, each subset of 90,000 samples.

Additionally, we adopt the weather transformation operations in [19] to simulate the natural scenarios. Specifically, there are four transformation operations (i.e., Snow, Frost, Fog, and Brightness), where each type has five levels of severity, resulting in a set $A$ of 20 fine-grained transformation operations. We generate a balanced-class dataset $T_{rank}$ with an equal size of the training set for each evaluation dataset by randomly selecting from training samples. We also use the benchmark dataset $T_{mark}$ provided by [19] to evaluate the effectiveness of defending subtle samples.

We select the ResNet20 [15] as the threat model for all evaluation datasets. Using different model architectures would introduce a new variable in the experiments. Therefore, we unify it to eliminate the impact. Specifically, the model is trained with augmentation of cropping and flipping for 200 epochs at an initial learning rate of 0.1, with a momentum multiplier of 0.9 weight decay with a multiplier of $5e^{-4}$, and batch size 128. The learning rate is cosine annealed to zero. The validation accuracy of the trained models for CIFAR10, CIFAR100, and CINIC10 are 92.58%, 69.09%, and 83.17%, respectively. The resultant models are used as the model under test M in our experiments.

Table 1 shows the statistics of samples under rank in our experiments. The second column shows the number of samples under rank in each dataset. The failing ratio describes the proportion of failing samples in the dataset. The subtle ratio shows the fraction of the failing samples with higher than $\theta$ confidence. The decreasing subtle ratios echo the challenges mentioned in Section 4.1.

### 5.4 Baselines

We aim to compare $A^3Rank$ with diverse categories of state-of-the-art techniques as baselines.

*DeepGini* [11] is the metric-based prioritization technique. It computes the gini impurity from the prediction vector and utilizes it as the metric for ranking.

*Monte Carlo Dropout* [12] (Dropout in short) is the uncertainty-based prioritization technique. It arbitrarily enables a dropout layer of the model under test in the inference time and performs $K$ stochastic forward passes over the sample under rank $s$ followed by averaging the variances. We configure the dropout rate of the dropout layer to 0.05 and the number of forward passes $K$ to 50.

*Dissector* [46] is the trend-based prioritization technique. Dissector trains a sequence of submodels for individual layers and measures the validity degree of confidence trend from the prediction vectors of these submodels. We configure to train the submodels, each with 5 epochs.

*LSA* [26] is the surprise-based prioritization technique. It trains a distribution on the intermediate output of a layer in the model under test over the whole training set for each class. It ranks the samples by the probability density to the distribution of its predicted class.

*Prima* [47] is the prioritization technique adopting the learning-to-rank concept. In their experiments, Prima generates 200 model mutants for the model under test and 100 sample mutants for each input sample. It then extracts their predefined features from the mutation results and learns a ranker model. We note that the original number of model mutants and sample mutants exceeds the affordable computation resource in our platform (i.e., 20000x inference time), so we have to reduce the number of model mutants to 20 and the number of sample mutants to 10.

Coverage-based techniques [29, 38] are not included because DeepGini has shown consistent and significant outperformance compared to them [11].

### 5.5 Experiment Settings

*5.5.1 Experiment 1.* We input the dataset $T_{rank}$ and its associated model to Alg. 1 and execute from line 1 to line 10 to produce a ranked list of samples $T^{\blacktriangleleft}_{rank}$ for $A^3Rank$. Using their algorithms, the baseline rankers also produce their ranked lists of samples $T^{\blacktriangleleft}_{rank}$. Then, we directly select the top-$\omega$ samples from each ranked list for measurement. We configure two labeling resources $\omega$: 1) $\omega$ is set to 10% of the dataset size, i.e., $\omega = 0.1 * |T_{rank}|$. We refer to this selection setting as TOP. 2) $\omega$ is set to the total number of failing samples in $T_{rank}$. (We note that the number of failing samples can be estimated via sampling in practice.) We refer to this selection setting as CUT. We configure three confidence-based rejectors Rs with thresholds $\theta$ of 0.7, 0.8, 0.9, respectively. For each rejector R, we further execute line 11 of Alg. 1 to produce a subset of subtle samples $T_{sub}$ that passes R for each set of resultant top-$\omega$ samples. As labeling each set of subtle samples $T_{sub}$ at line 11 of Alg. 1 requires an external process , we assure the labeling quality is perfect in the evaluation. We measure the number of discovered subtle samples and the throughput ratio of subtle samples, i.e., the number of subtle samples over the number of failing samples. The result is averaged over 10 runs.

*5.5.2 Experiment 2.* We use $T_{sub}$ in Experiment 1 to construct a NIC model as the detector D. More specifically, the NIC algorithm accepts a set of samples and a DL model and then produces a model containing a sequence of one-class SVM (OSVM) models and an OSVM-based predictor. We input $T_{sub}$ and the DL model under test M to the NIC algorithm and acquire the resultant NIC model as D. Given a novel sample, we input it into the NIC model to predict whether it is in distribution to $T_{sub}$, and if so, the detector rejects this sample. We input all failing samples that passed the rejector in the benchmark dataset $T_{mark}$ to the trained D and measure the ratio of the number of samples rejected by D over these failing samples. This ratio is also known as the *defense success rate*.

*5.5.3 Experiment 3.* We also investigate $A^3Rank$ to rank the whole test dataset to cover the detection of all failing samples compared with state-of-the-art TCP techniques. The effectiveness is evaluated by the improvement ratio over the ranking by the Random technique. We measure the ratio of the number of failing samples in the top-$\omega$ samples as in Experiment 1 and report the improvement ratio.

*5.5.4 Experiment 4 (Ablation Study).* We ablate each alignment (namely $g_1$, $g_2$, and $g_3$) of $A^3Rank$ to produce an ablated version of $A^3Rank$. We denote the ablated versions of $A^3Rank$ without $g_1$, $g_2$, and $g_3$ by $A^3R_{ank}^{-g_1}$, $A^3R_{ank}^{-g_2}$, $A^3R_{ank}^{-g_3}$, respectively. We repeat Experiment 1 for each ablated version and measure the number of discovered subtle samples.

## 5.6  Experiment Results

Table 2 summarizes the results of Experiment 1 under different rejection settings for R. The results of LSA and Prima are omitted due to the limited effectiveness in finding failing samples (see Table 4) for space consideration. In each cell, there are two values. The one at the top is the discovered number of subtle samples, and the other at the bottom is the throughput ratio. The cell with the most discovered subtle samples in each column is highlighted in bold.

Across the board, $A^3Rank$ consistently detects more subtle samples compared with the baselines under all the settings, and the throughput ratio of subtle samples is also larger than the baselines, except for just one case of CUT with $\theta = 7$ on the CIFAR100 dataset. We note that under that case among the effective techniques of Dropout, Dissector, and $A^3Rank$, the maximum difference is less than 5% (i.e., $10.71/10.31 - 1 = 3.88\%$), not to mention the difference between $A^3Rank$ and Dropout is only $1.03\% = 10.71/10.60 - 1$.

The differences between $A^3Rank$ and the best baseline results are large. Taking the average of the same rejection threshold, $A^3Rank$ outperforms the largest of the baselines by 52.78%, 103.93%, and 301.52% on the thresholds of $\theta = 0.7$, $\theta = 0.8$, and $\theta = 0.9$, respectively, which are significant. Noticeably, the differences across the increasing $\theta$ are increasing, indicating that $A^3Rank$ is more effective in finding the failing samples with higher prediction confidences.

Regarding the throughput ratios, $A^3Rank$ also achieves significantly more effective results than the best baselines. Similarly, on account of the average under the same rejection threshold, $A^3Rank$ outperforms the largest of the baselines by 24.73%, 217.07%, and 249.11% (163.63% averagely) on $\theta = \{0.7, 0.8, 0.9\}$, respectively. The increasing differences are consistent with the differences in the number of subtle samples, value-adding on top of the effectiveness of $A^3Rank$ in ranking these challenging subtle samples.

> **Answering RQ1**
>
> $A^3Rank$ consistently and significantly outperforms the baselines in finding subtle samples up to 301.52%, and the ranking effectiveness increases with the tasks to discovering failing samples with higher confidence.

Table 3 summarizes the results of the defense success rates for each technique on each setting using our $A^3Rank$ framework. Note that the NIC algorithm requires sufficient samples to produce an operable OSVM model. In the table, if the size of discovered subtle samples is insufficient for applying the NIC algorithm or the algorithm cannot find a supportive smallest hypersphere for the

Table 2. Results of Prioritization for Subtle Samples

| Technique | CIFAR10 | | | | | | CIFAR100 | | | | | | CINIC10 | | | | | |
|---|---|---|---|---|---|---|---|---|---|---|---|---|---|---|---|---|---|---|
| | TOP | | | CUT | | | TOP | | | CUT | | | TOP | | | CUT | | |
| | θ=0.7 | θ=0.8 | θ=0.9 | θ=0.7 | θ=0.8 | θ=0.9 | θ=0.7 | θ=0.8 | θ=0.9 | θ=0.7 | θ=0.8 | θ=0.9 | θ=0.7 | θ=0.8 | θ=0.9 | θ=0.7 | θ=0.8 | θ=0.9 |
| DeepGini | 84<br>2.78% | 0<br>0.00% | 0<br>0.00% | 1012<br>25.09% | 113<br>2.80% | 0<br>0.00% | 0<br>0.00% | 0<br>0.00% | 0<br>0.00% | 0<br>0.00% | 0<br>0.00% | 0<br>0.00% | 0<br>0.00% | 0<br>0.00% | 0<br>0.00% | 59<br>0.66% | 0<br>0.00% | 0<br>0.00% |
| Dropout | 865<br>30.38% | 379<br>13.31% | 82<br>2.88% | 1379<br>34.29% | 720<br>17.9% | 186<br>4.62% | 195<br>7.01% | 37<br>1.33% | 4<br>0.14% | 1194<br>10.71% | 329<br>2.95% | 31<br>0.28% | 681<br>16.14% | 178<br>4.22% | 19<br>0.45% | 1423<br>18.82% | 447<br>5.91% | 55<br>0.73% |
| Dissector | 877<br>27.39% | 424<br>13.24% | 119<br>3.72% | 1360<br>34.98% | 802<br>20.63% | 358<br>9.21% | 229<br>5.07% | 71<br>1.57% | 8<br>0.18% | 1386<br>10.31% | 416<br>3.31% | 40<br>0.30% | 478<br>8.07% | 109<br>1.84% | 4<br>0.07% | 1571<br>17.56% | 647<br>7.23% | 153<br>1.71% |
| **A³Rank** | **1174<br>34.49%** | **664<br>19.51%** | **270<br>7.93%** | **1844<br>40.72%** | **1182<br>26.10%** | **563<br>12.43%** | **471<br>11.19%** | **176<br>4.18%** | **37<br>0.88%** | **1429<br>10.60%** | **611<br>4.53%** | **157<br>1.16%** | **1298<br>21.34%** | **606<br>9.96%** | **172<br>2.83%** | **2352<br>24.02%** | **1193<br>12.18%** | **405<br>4.14%** |

Table 3. Results of Defense for Subtle Samples

| Technique | CIFAR10 | | | | | | CIFAR100 | | | | | | CINIC10 | | | | | |
|---|---|---|---|---|---|---|---|---|---|---|---|---|---|---|---|---|---|---|
| | TOP | | | CUT | | | TOP | | | CUT | | | TOP | | | CUT | | |
| | θ=0.7 | θ=0.8 | θ=0.9 | θ=0.7 | θ=0.8 | θ=0.9 | θ=0.7 | θ=0.8 | θ=0.9 | θ=0.7 | θ=0.8 | θ=0.9 | θ=0.7 | θ=0.8 | θ=0.9 | θ=0.7 | θ=0.8 | θ=0.9 |
| DeepGini | ⊖ | ⊖ | 55.98% | ⊖ | ⊖ | ⊖ | ⊖ | ⊖ | ⊖ | ⊖ | ⊖ | ⊖ | ⊖ | ⊖ | ⊖ | ⊖ | ⊖ | ⊖ |
| Dropout | 62.13% | 61.25% | 75.12% | 57.28% | 59.58% | 58.97% | ⊖ | ⊖ | ⊖ | 38.17% | 26.09% | ⊖ | 58.01% | 25.21% | ⊖ | 64.50% | 60.17% | ⊖ |
| Dissector | 55.27% | 64.58% | 78.33% | 33.65% | 41.25% | 43.29% | ⊖ | ⊖ | ⊖ | 53.47% | 76.83% | ⊖ | 20.90% | ⊖ | ⊖ | 65.92% | 63.61% | 30.18% |
| **A³Rank** | **67.50%** | **75.73%** | **93.29%** | **64.40%** | **72.50%** | **72.56%** | ⊖ | ⊖ | **99.90%** | **61.03%** | **85.63%** | **99.90%** | **55.58%** | **75.93%** | ⊖ | **71.11%** | **74.98%** | **33.33%** |

Note: ⊖ = The NIC algorithm cannot create a detector due to not enough data.

Table 4. Results of Improvement Ratios for Detecting Failing Samples over Random Technique

| Technique | CIFAR10 | | CIFAR100 | | CINIC10 | |
|---|---|---|---|---|---|---|
| | TOP | CUT | TOP | CUT | TOP | CUT |
| DeepGini | 4.11x | 3.73x | 2.13x | 1.78x | 3.43x | 3.08x |
| Dropout | 3.87x | 3.72x | 1.46x | 1.54x | 2.60x | 2.59x |
| Dissector | 4.36x | 3.60x | 2.11x | 1.86x | 3.65x | 3.07x |
| LSA | 2.38x | 2.26x | 1.73x | 1.51x | 1.71x | 1.58x |
| Prima | 2.56x | 2.48x | 1.76x | 1.55x | 2.31x | 1.87x |
| $A^3Rank$ | **4.63x** | **4.19x** | **2.21x** | **1.86x** | **3.75x** | **3.35x** |

OSVM model, we mark the cell with ⊖ to indicate that there is a failure instance of the technique in the experiment. We highlight the largest results, which exceed other techniques in 5% differences.

Overall speaking, $A^3Rank$ framework is able to generate effective detectors by utilizing the discovered subtle samples to defend subtle samples during the inference time. For instance, there are 16 effective detectors over 18 settings (88.88%) using the subtle samples discovered by $A^3Rank$. The other two cases are due to insufficiently discovered subtle samples, which can be addressed by capturing more natural samples for ranking in real-world applications. All baselines experienced significantly higher failure rates in generating detectors.

Under the settings of the same number of samples to rank, the detectors produced with the samples revealed by $A^3Rank$ are generally more effective than those revealed by the baselines. Recall the results in Table 2. Observably, the more subtle samples are discovered, the more effective the detectors are produced. We find that DeepGini can hardly produce effective detectors. Therefore, we skip it for comparison. Compared with the larger results of Dropout and Dissector where applicable, regarding the defense success rate where there are more than 5% differences, $A^3Rank$ enlarges the performance with 5.19% to 7.57%, 8.80% to 50.72%, and 3.15% to 14.96% on $\theta = \{0.7, 0.8, 0.9\}$, respectively. The average improvement across three $\theta$ settings is 12.67%, which is large.

> **Answering RQ2**
>
> $A^3Rank$ framework is effective in producing detectors to enhance the application models, and the detectors produced with the subtle samples discovered by $A^3Rank$ are more effective than those by the baselines.

Table 4 summarizes the results of Experiment 3 achieved by the baselines and our technique $A^3Rank$ to rank the whole test dataset. The second to the last columns show the results of the improvement ratios of detecting failing samples in the two selection settings over the Random technique. The results of the techniques in each setting that achieves the highest improvement ratio are shown in bold.

Overall speaking, $A^3Rank$ consistently achieves the highest improvement ratios in all experimental settings. Among the baselines, LSA and Prima are less effective in discovering the failing samples compared with the other techniques, and DeepGini and Dissector achieve relatively high performance and outperform the other three baselines significantly, except in the CUT column on the CIFAR10 dataset. DeepGini and Dissector each achieve better results than the other in three cases, showing that the previous state-of-the-art techniques are not consistent in outperforming one another. Nonetheless, they are still less effective than $A^3Rank$.

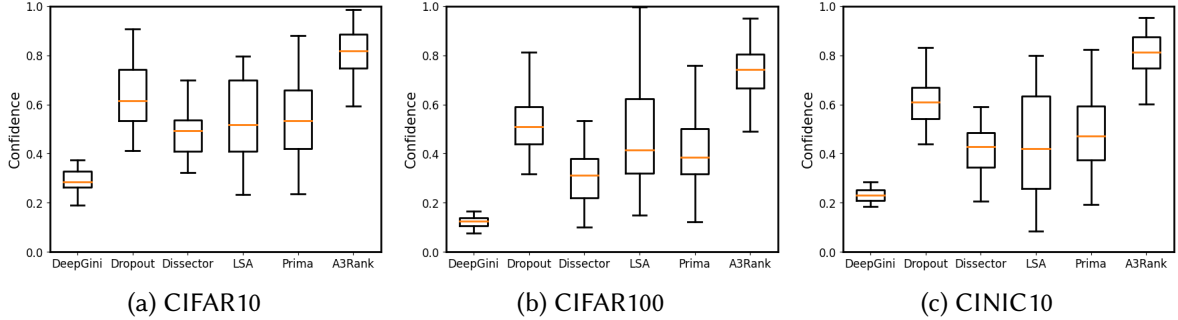

Fig. 5.  Confidence distribution of top 50 failing samples in the ranked list.

We conduct the *Wilcoxon Signed Rank Test* [39] (a paired test) on each pair of lists of detection ratios between $A^3Rank$ and each baseline. All the tests show that the respective two results are significantly different at the 5% significance level.

Fig. 5 visualizes the confidence distribution of the top-50 failing samples in the resultant ranked list of each experimental setting to demonstrate their preference in ranking samples, where the number of 50 is arbitrarily selected equaling to 1‰ of training samples in CIFAR datasets.

The figure shows that $A^3Rank$ exposes the failing samples with higher prediction confidences than the baselines, and it is consistent across the three datasets. The most interesting finding is that DeepGini performs significantly ineffective compared with $A^3Rank$. DeepGini also adopts prediction confidence to formulate its metric for ranking. $A^3Rank$ designs three alignment terms to apply to the vanilla prediction confidence and consistently achieves the best results, which validates the novelty of $A^3Rank$ in prioritizing failing samples with higher prediction confidences.

> **Answering RQ3**
>
> $A^3Rank$ consistently outperforms the baselines in detecting failing samples. Its top-ranked samples are with higher prediction confidence than the baselines.

Fig. 6 presents the differences between $A^3Rank$ and its ablated versions without each regulation of $g_1$, $g_2$, and $g_3$ (denoted by $A^3R_{ank}^{-g1}$, $A^3R_{ank}^{-g2}$, and $A^3R_{ank}^{-g3}$), respectively. In each subfigure, the bars sequentially present their results, where the y-axis is the number of discovered subtle samples.

It is observed that, in general, eliminating any alignment term decreases the effectiveness of $A^3Rank$. The $g_1$ alignment dramatically disturbs the effectiveness of $A^3Rank$ significantly. The $g_2$ alignment slightly affects the effectiveness of $A^3Rank$. While the effects of $g_3$ alignment lie between the effects of $g_1$ and $g_2$. The complete version of $A^3Rank$ with all three alignment terms achieves the overall highest results.

> **Answering RQ4**
>
> Each alignment term in $A^3Score$ (Eq. 6) contributes positively to the effectiveness of $A^3Rank$.

## 5.7 Threats to Validity

A threat to validity lies in the concerns about the robustness improvement ability of $A^3Rank$. To reduce this threat, we conduct experiments to mimic the real-world scenario with 5 rounds of data collection. For each round, we generate a $T_{rank}$ and repeat the CUT procedure of Experiment 3 on CIFAR10 to find the top-$\omega$ failing samples along with the discovered failing samples in former rounds for retraining. Fig. 7 shows the results, where the left y-axis is robust accuracy measured on $T_{mark}$ and the right y-axis is the accumulated discovered failing samples. We observe that with more rounds of collected data to rank, the robust accuracy increases with more failing samples discovered. Besides, the experiments only evaluate one model architecture. We note that different model architectures perform similarly on the benchmarks[1] by PyTorch. Another threat to validity is the imprecise decision of the detector D. We are aware that the detectors generated by the NIC algorithm make false positive decisions. However, the false positive rate is small according to their paper [31], and the survey [17] outlines that the true negative predictions are tolerable in models with reject options and could be deferred to a human expert in certain application scenarios. The implementation may contain bugs unknown to us. We have tested that the downloaded code is reproducible. The experiments are limited to one kind of rejector and the chosen hyperparameters. Using more detection algorithms, datasets, DL models, rejectors, and hyperparameters can generalize the result, and we leave the generalization in future work.

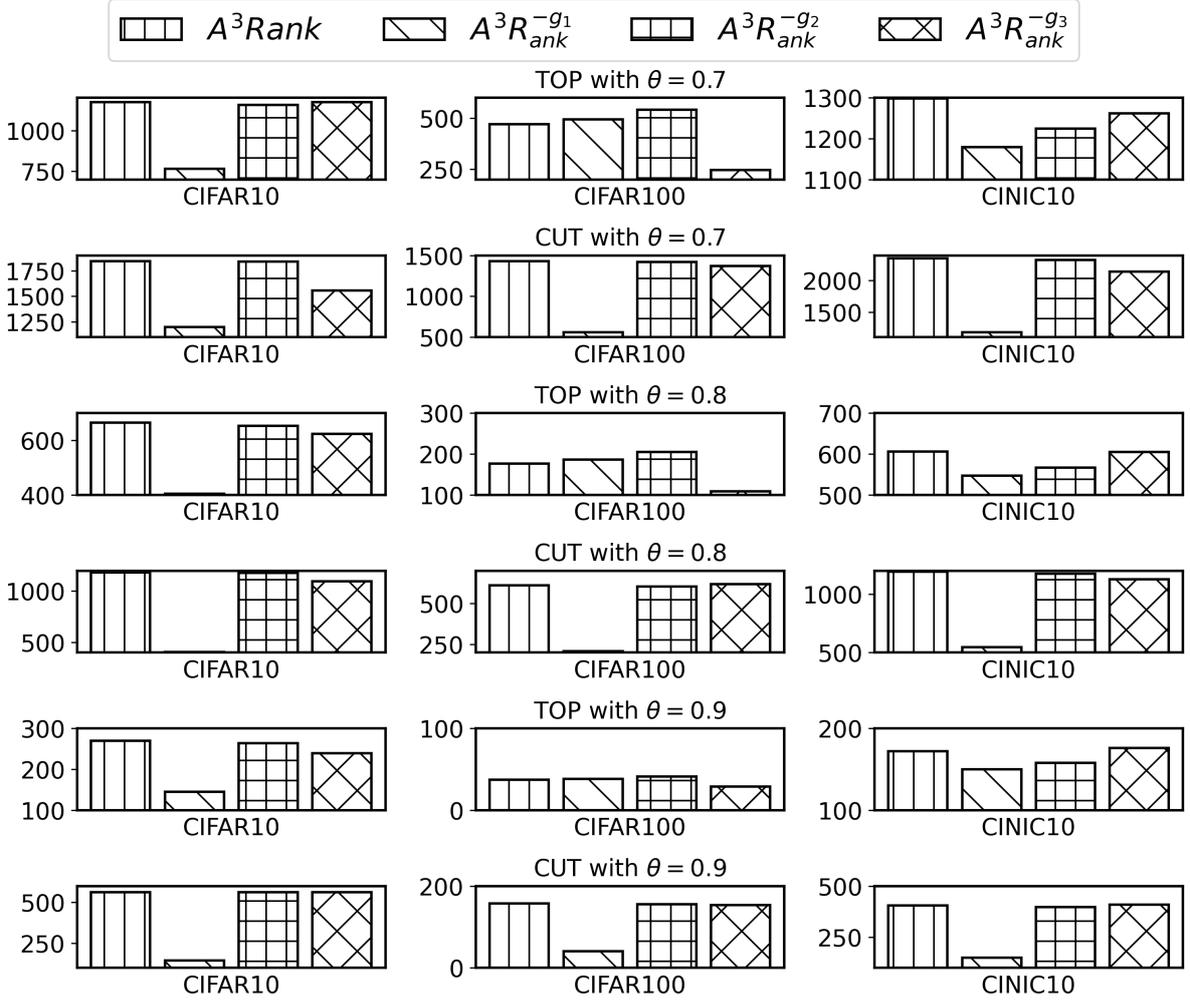

Fig. 6. Results of the ablated alignment terms of $A^3Rank$

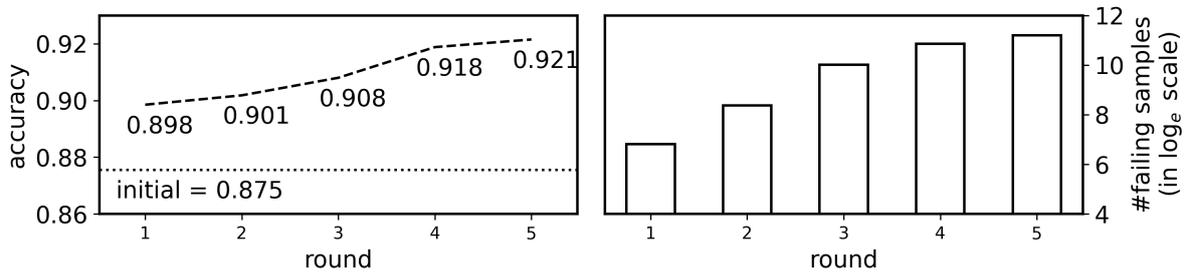

Fig. 7. Results of robustness improvement of $A^3Rank$.

## 6 RELATED WORK

Test case prioritization and selection techniques [23, 24, 28, 30, 50, 53] have been intensively studied. DeepGini [11], Dissector[46], LSA [26], Prima [47] and Dropout [12] hae been reviewed and compared in Section 5. Test case selection can involve choosing a subset of representative cases for more efficient testing rather than testing the entire set [23, 28, 53]. Our work effectively finds the subtle samples and may have the potential to complement the reduced test set produced by these works for evaluation. Other works [24, 30, 50] identify the samples triggering abnormal neuron activations within the model under test. Our work identifies the failing sample with higher prediction confidence and could potentially ease profiling by providing subtle examples for analysis of the neuron activation patterns.

A range of related works aims to incorporate a reject option for machine learning and deep learning models [3, 13, 21, 22, 37, 40, 41, 44]. Many techniques focus on novelty (outlier) rejection [37, 40, 44]. It has been reported that the outliers are much easier to detect via accuracy thresholding [37], support vector machine [40], and uncertainty estimation [44]. Another technique category is the ambiguity rejection with different model architectures [3, 13, 21, 22, 41]. Some rejectors are constructed independently of the classifiers used, but this often results in sub-optimal rejection performance [22]. Other approaches leverage classifier information to build rejectors [3, 21], and some further attempts to integrate rejectors with classifiers [13, 41]. Our work falls into this latter category and can be used to identify subtle samples for better integration with these techniques.

Another related line of work focuses on adversarial example detection and anomaly detection [10, 14, 31, 33–35, 48]. These techniques can be further classified into supervised detection [10, 14, 33, 35] and unsupervised detection [31, 34, 48] methods. The feature-based approaches [14, 35] use clean and adversarial examples to train their detectors. The statistical-based approaches[10, 33] perform a statistical measurement to cut the separation between these two sets of samples. On the other hand, unsupervised detectors propose feature squeezing [48], denoising [34], and invariant [31] to produce effective detectors. Our work adopts the state-of-the-art unsupervised method [31] to generate a detector. Our work may help to find subtle samples along with clean samples with higher confidence to construct an improved supervised detector.

## 7 CONCLUSION

We have presented $A^3Rank$, a novel and effective prioritization technique for ranking samples of deep learning models with a reject option. $A^3Rank$ comes with a novel augmentation alignment analysis to score the samples under rank. The experiment results have shown that $A^3Rank$ is significantly more effective in ranking failing samples escaping the confidence-based rejectors and detecting these samples during inference time compared with the baselines. Owing to revealing more such samples, $A^3Rank$ has been shown to be able to build more defenders than the baselines.

---

[1]https://pytorch.org/vision/stable/models.html#classification

## REFERENCES

[1] Ahmed Aldahdooh, Wassim Hamidouche, and Olivier Déforges. 2023. Revisiting model's uncertainty and confidences for adversarial example detection. *Applied Intelligence* 53, 1 (01 Jan 2023), 509–531. https://doi.org/10.1007/s10489-022-03373-y

[2] David Berend, Xiaofei Xie, Lei Ma, Lingjun Zhou, Yang Liu, Chi Xu, and Jianjun Zhao. 2020. Cats Are Not Fish: Deep Learning Testing Calls for Out-Of-Distribution Awareness. In *2020 35th IEEE/ACM International Conference on Automated Software Engineering (ASE)*. 1041–1052.

[3] Abdenour Bounsiar, Edith Grall, and Pierre Beauseroy. 2007. A Kernel Based Rejection Method for Supervised Classification. *World Academy of Science, Engineering and Technology, International Journal of Computer, Electrical, Automation, Control and Information Engineering* 1 (2007), 3907–3916.

[4] Nicholas Carlini and David Wagner. 2017. Adversarial Examples Are Not Easily Detected: Bypassing Ten Detection Methods. In *Proceedings of the 10th ACM Workshop on Artificial Intelligence and Security* (Dallas, Texas, USA) *(AISec '17)*. Association for Computing Machinery, New York, NY, USA, 3–14. https://doi.org/10.1145/3128572.3140444

[5] chenyaofo. 2023. Implementation of pytorch-cifar-models. https://github.com/chenyaofo/pytorch-cifar-models.

[6] Charles Corbière, Nicolas Thome, Avner Bar-Hen, Matthieu Cord, and Patrick Pérez. 2019. *Addressing Failure Prediction by Learning Model Confidence.* Curran Associates Inc., Red Hook, NY, USA.

[7] Luke N. Darlow, Elliot J. Crowley, Antreas Antoniou, and Amos J. Storkey. 2018. CINIC-10 is not ImageNet or CIFAR-10. arXiv:1810.03505 [cs.CV]

[8] Swaroopa Dola, Matthew B. Dwyer, and Mary Lou Soffa. 2021. Distribution-Aware Testing of Neural Networks Using Generative Models. In *Proceedings of the 43rd International Conference on Software Engineering* (Madrid, Spain) *(ICSE '21)*. IEEE Press, 226–237. https://doi.org/10.1109/ICSE43902.2021.00032

[9] Isaac Dunn, Hadrien Pouget, Daniel Kroening, and Tom Melham. 2021. Exposing Previously Undetectable Faults in Deep Neural Networks. In *Proceedings of the 30th ACM SIGSOFT International Symposium on Software Testing and Analysis* (Virtual, Denmark) *(ISSTA 2021)*. Association for Computing Machinery, New York, NY, USA, 56–66. https://doi.org/10.1145/3460319.3464801

[10] Reuben Feinman, Ryan R. Curtin, Saurabh Shintre, and Andrew B. Gardner. 2017. Detecting Adversarial Samples from Artifacts. arXiv:1703.00410 [stat.ML]

[11] Yang Feng, Qingkai Shi, Xinyu Gao, Jun Wan, Chunrong Fang, and Zhenyu Chen. 2020. DeepGini: Prioritizing Massive Tests to Enhance the Robustness of Deep Neural Networks. In *Proceedings of the 29th ACM SIGSOFT International Symposium on Software Testing and Analysis* (Virtual Event, USA) *(ISSTA 2020)*. Association for Computing Machinery, New York, NY, USA, 177–188. https://doi.org/10.1145/3395363.3397357

[12] Yarin Gal and Zoubin Ghahramani. 2016. Dropout as a Bayesian Approximation: Representing Model Uncertainty in Deep Learning. In *Proceedings of the 33rd International Conference on International Conference on Machine Learning - Volume 48* (New York, NY, USA) *(ICML'16)*. JMLR.org, 1050–1059.

[13] Yonatan Geifman and Ran El-Yaniv. 2019. Selectivenet: A deep neural network with an integrated reject option. In *International conference on machine learning*. PMLR, 2151–2159.

[14] Kathrin Grosse, Praveen Manoharan, Nicolas Papernot, Michael Backes, and Patrick McDaniel. 2017. On the (Statistical) Detection of Adversarial Examples. arXiv:1702.06280 [cs.CR]

[15] Kaiming He, Xiangyu Zhang, Shaoqing Ren, and Jian Sun. 2015. Deep Residual Learning for Image Recognition. arXiv:1512.03385 [cs.CV]

[16] Simon Hecker, Dengxin Dai, and Luc Van Gool. 2018. Failure Prediction for Autonomous Driving. In *2018 IEEE Intelligent Vehicles Symposium (IV)*. 1792–1799. https://doi.org/10.1109/IVS.2018.8500495

[17] Kilian Hendrickx, Lorenzo Perini, Dries Van der Plas, Wannes Meert, and Jesse Davis. 2021. Machine Learning with a Reject Option: A survey. arXiv:2107.11277 [cs.LG]

[18] Dan Hendrycks and Thomas Dietterich. 2019. Benchmarking Neural Network Robustness to Common Corruptions and Perturbations. *Proceedings of the International Conference on Learning Representations* (2019).

[19] Dan Hendrycks and Thomas Dietterich. 2019. Benchmarking Neural Network Robustness to Common Corruptions and Perturbations. In *International Conference on Learning Representations*. https://openreview.net/forum?id=HJz6tiCqYm

[20] Dan Hendrycks and Kevin Gimpel. 2017. A Baseline for Detecting Misclassified and Out-of-Distribution Examples in Neural Networks. In *International Conference on Learning Representations*. https://openreview.net/forum?id=Hkg4TI9xl

[21] Jay Heo, Hae Beom Lee, Saehoon Kim, Juho Lee, Kwang Joon Kim, Eunho Yang, and Sung Ju Hwang. 2018. Uncertainty-Aware Attention for Reliable Interpretation and Prediction. In *Proceedings of the 32nd International Conference on Neural Information Processing Systems* (Montréal, Canada) *(NIPS'18)*. Curran Associates Inc., Red Hook, NY, USA.

[22] Wladyslaw Homenda, Marcin Luckner, and Witold Pedrycz. 2014. Classification with rejection based on various SVM techniques. In *2014 International Joint Conference on Neural Networks (IJCNN)*. 3480–3487. https://doi.org/10.1109/IJCNN.2014.6889655

[23] Qiang Hu, Yuejun Guo, Xiaofei Xie, Maxime Cordy, Mike Papadakis, Lei Ma, and Yves Le Traon. 2023. Aries: Efficient Testing of Deep Neural Networks via Labeling-Free Accuracy Estimation. In *Proceedings of the 45rd International Conference on Software Engineering* (Melbourne, Australia) *(ICSE '23)*. https://arxiv.org/abs/2207.10942

[24] Zhenlan Ji, Pingchuan Ma, Yuanyuan Yuan, and Shuai Wang. 2023. CC: Causality-Aware Coverage Criterion for Deep Neural Networks. In *Proceedings of the 45rd International Conference on Software Engineering* (Melbourne, Australia) *(ICSE '23)*.

[25] Xiaoqian Jiang, Melanie Osl, Jihoon Kim, and Lucila Ohno-Machado. 2011. Calibrating predictive model estimates to support personalized medicine. *Journal of the American Medical Informatics Association* 19, 2 (10 2011), 263–274. https://doi.org/10.1136/amiajnl-2011-000291

[26] Jinhan Kim, Robert Feldt, and Shin Yoo. 2019. Guiding Deep Learning System Testing Using Surprise Adequacy. In *Proceedings of the 41st International Conference on Software Engineering* (Montreal, Quebec, Canada) *(ICSE '19)*. IEEE Press, 1039–1049. https://doi.org/10.1109/ICSE.2019.00108

[27] Alex Krizhevsky, Geoffrey Hinton, et al. 2009. Learning multiple layers of features from tiny images. *Master's thesis, Department of Computer Science, University of Toronto* (2009).

[28] Zijie Li, Long Zhang, Jun Yan, Jian Zhang, Zhenyu Zhang, and T. H. Tse. 2020. PEACEPACT: Prioritizing Examples to Accelerate Perturbation-Based Adversary Generation for DNN Classification Testing. In *2020 IEEE 20th International Conference on Software Quality, Reliability and Security (QRS)*. 406–413. https://doi.org/10.1109/QRS51102.2020.00059

[29] Lei Ma, Felix Juefei-Xu, Fuyuan Zhang, Jiyuan Sun, Minhui Xue, Bo Li, Chunyang Chen, Ting Su, Li Li, Yang Liu, Jianjun Zhao, and Yadong Wang. 2018. DeepGauge: Multi-Granularity Testing Criteria for Deep Learning Systems. In *Proceedings of the 33rd ACM/IEEE International Conference on Automated Software Engineering* (Montpellier, France) *(ASE '18)*. Association for Computing Machinery, New York, NY, USA, 120–131. https://doi.org/10.1145/3238147.3238202

[30] Shiqing Ma, Yingqi Liu, Wen-Chuan Lee, Xiangyu Zhang, and Ananth Grama. 2018. MODE: Automated Neural Network Model Debugging via State Differential Analysis and Input Selection. In *Proceedings of the 2018 26th ACM Joint Meeting on European Software Engineering Conference and Symposium on the Foundations of Software Engineering* (Lake Buena Vista, FL, USA) *(ESEC/FSE 2018)*. Association for Computing Machinery, New York, NY, USA, 175–186. https://doi.org/10.1145/3236024.3236082

[31] Shiqing Ma, Yingqi Liu, Guanhong Tao, Wen-Chuan Lee, and Xiangyu Zhang. 2019. Nic: Detecting adversarial samples with neural network invariant checking. In *26th Annual Network And Distributed System Security Symposium (NDSS 2019)*. Internet Soc.

[32] Wei Ma, Mike Papadakis, Anestis Tsakmalis, Maxime Cordy, and Yves Le Traon. 2021. Test Selection for Deep Learning Systems. *ACM Trans. Softw. Eng. Methodol.* 30, 2, Article 13 (jan 2021), 22 pages. https://doi.org/10.1145/3417330

[33] Xingjun Ma, Bo Li, Yisen Wang, Sarah M. Erfani, Sudanthi Wijewickrema, Grant Schoenebeck, Dawn Song, Michael E. Houle, and James Bailey. 2018. Characterizing Adversarial Subspaces Using Local Intrinsic Dimensionality. arXiv:1801.02613 [cs.LG]

[34] Dongyu Meng and Hao Chen. 2017. MagNet: A Two-Pronged Defense against Adversarial Examples. In *Proceedings of the 2017 ACM SIGSAC Conference on Computer and Communications Security* (Dallas, Texas, USA) *(CCS '17)*. Association for Computing Machinery, New York, NY, USA, 135–147. https://doi.org/10.1145/3133956.3134057

[35] Jan Hendrik Metzen, Tim Genewein, Volker Fischer, and Bastian Bischoff. 2017. On Detecting Adversarial Perturbations. In *International Conference on Learning Representations*. https://openreview.net/forum?id=SJzCSf9xg

[36] Mary M. Moya and Don R. Hush. 1996. Network constraints and multi-objective optimization for one-class classification. *Neural Networks* 9, 3 (1996), 463–474. https://doi.org/10.1016/0893-6080(95)00120-4

[37] Malik Sajjad Ahmed Nadeem, Jean-Daniel Zucker, and Blaise Hanczar. 2009. Accuracy-Rejection Curves (ARCs) for Comparing Classification Methods with a Reject Option. In *Proceedings of the third International Workshop on Machine Learning in Systems Biology (Proceedings of Machine Learning Research, Vol. 8)*, Sašo Džeroski, Pierre Guerts, and Juho Rousu (Eds.). PMLR, Ljubljana, Slovenia, 65–81. https://proceedings.mlr.press/v8/nadeem10a.html

[38] Kexin Pei, Yinzhi Cao, Junfeng Yang, and Suman Jana. 2019. DeepXplore: Automated Whitebox Testing of Deep Learning Systems. *Commun. ACM* 62, 11 (oct 2019), 137–145. https://doi.org/10.1145/3361566

[39] Denise Rey and Markus Neuhäuser. 2011. *Wilcoxon-Signed-Rank Test*. Springer Berlin Heidelberg, Berlin, Heidelberg, 1658–1659. https://doi.org/10.1007/978-3-642-04898-2_616

[40] Bernhard Schölkopf, Robert Williamson, Alex Smola, John Shawe-Taylor, and John Platt. 1999. Support Vector Method for Novelty Detection. In *Proceedings of the 12th International Conference on Neural Information Processing Systems* (Denver, CO) *(NIPS'99)*. MIT Press, Cambridge, MA, USA, 582–588.

[41] Angelo Sotgiu, Ambra Demontis, Marco Melis, Battista Biggio, Giorgio Fumera, Xiaoyi Feng, and Fabio Roli. 2020. Deep neural rejection against adversarial examples. *EURASIP Journal on Information Security* 2020, 1 (07 Apr 2020), 5.

https://doi.org/10.1186/s13635-020-00105-y

[42] Yiyou Sun, Chuan Guo, and Yixuan Li. 2021. ReAct: Out-of-distribution Detection With Rectified Activations. In *Advances in Neural Information Processing Systems*, A. Beygelzimer, Y. Dauphin, P. Liang, and J. Wortman Vaughan (Eds.). https://openreview.net/forum?id=IBVBtz_sRSm

[43] Luke Taylor and Geoff Nitschke. 2018. Improving Deep Learning with Generic Data Augmentation. In *2018 IEEE Symposium Series on Computational Intelligence (SSCI)*. 1542–1547. https://doi.org/10.1109/SSCI.2018.8628742

[44] Dennis Ulmer and Giovanni Cinà. 2021. Know your limits: Uncertainty estimation with relu classifiers fails at reliable ood detection. In *Uncertainty in Artificial Intelligence*. PMLR, 1766–1776.

[45] Kush R. Varshney. 2011. A risk bound for ensemble classification with a reject option. In *2011 IEEE Statistical Signal Processing Workshop (SSP)*. 769–772. https://doi.org/10.1109/SSP.2011.5967817

[46] Huiyan Wang, Jingwei Xu, Chang Xu, Xiaoxing Ma, and Jian Lu. 2020. DISSECTOR: Input Validation for Deep Learning Applications by Crossing-layer Dissection. In *2020 IEEE/ACM 42nd International Conference on Software Engineering (ICSE)*. 727–738.

[47] Zan Wang, Hanmo You, Junjie Chen, Yingyi Zhang, Xuyuan Dong, and Wenbin Zhang. 2021. Prioritizing Test Inputs for Deep Neural Networks via Mutation Analysis. In *2021 IEEE/ACM 43rd International Conference on Software Engineering (ICSE)*. 397–409. https://doi.org/10.1109/ICSE43902.2021.00046

[48] Weilin Xu, David Evans, and Yanjun Qi. 2018. Feature Squeezing: Detecting Adversarial Examples in Deep Neural Networks. In *Proceedings 2018 Network and Distributed System Security Symposium*. Internet Society. https://doi.org/10.14722/ndss.2018.23198

[49] Yuanyuan Yuan, Qi Pang, and Shuai Wang. 2023. Revisiting Neuron Coverage for DNN Testing: A Layer-Wise and Distribution-Aware Criterion. In *2023 IEEE/ACM 45th International Conference on Software Engineering (ICSE)*. 1200–1212. https://doi.org/10.1109/ICSE48619.2023.00107

[50] Kai Zhang, Yongtai Zhang, Liwei Zhang, Hongyu Gao, Rongjie Yan, and Jun Yan. 2020. Neuron Activation Frequency Based Test Case Prioritization. In *2020 International Symposium on Theoretical Aspects of Software Engineering (TASE)*. 81–88. https://doi.org/10.1109/TASE49443.2020.00020

[51] Mengshi Zhang, Yuqun Zhang, Lingming Zhang, Cong Liu, and Sarfraz Khurshid. 2018. DeepRoad: GAN-Based Metamorphic Testing and Input Validation Framework for Autonomous Driving Systems. In *Proceedings of the 33rd ACM/IEEE International Conference on Automated Software Engineering* (Montpellier, France) *(ASE '18)*. Association for Computing Machinery, New York, NY, USA, 132–142. https://doi.org/10.1145/3238147.3238187

[52] Ziyuan Zhong, Yuchi Tian, and Baishakhi Ray. 2021. Understanding Local Robustness of Deep Neural Networks under Natural Variations. In *Fundamental Approaches to Software Engineering*, Esther Guerra and Mariëlle Stoelinga (Eds.). Springer International Publishing, Cham, 313–337.

[53] Jianyi Zhou, Feng Li, Jinhao Dong, Hongyu Zhang, and Dan Hao. 2020. Cost-Effective Testing of a Deep Learning Model through Input Reduction. In *2020 IEEE 31st International Symposium on Software Reliability Engineering (ISSRE)*. 289–300. https://doi.org/10.1109/ISSRE5003.2020.00035

[54] Fei Zhu, Zhen Cheng, Xu-Yao Zhang, and Cheng-Lin Liu. 2022. Rethinking Confidence Calibration for Failure Prediction. In *Computer Vision – ECCV 2022*, Shai Avidan, Gabriel Brostow, Moustapha Cissé, Giovanni Maria Farinella, and Tal Hassner (Eds.). Springer Nature Switzerland, Cham, 518–536.